\documentclass[aps,pra,superscriptaddress,floatfix,showpacs,10pt, twocolumn]{revtex4}
\usepackage[dvips]{graphicx}
\usepackage{amsmath,amsfonts,amssymb,graphics,graphicx,epsfig,color,times,bbm}

\begin{document}

\title{Generation of cluster states}
\author{Ping Dong\footnote{pingdong@ahu.edu.cn}}

\author{Zheng-Yuan Xue}
\author{Ming Yang}
\author{Zhuo-Liang Cao\footnote{zlcao@ahu.edu.cn(Corresponding~Author)}}

\affiliation{School of Physics {\&} Material Science, Anhui
University, Hefei, 230039, P R China}

\pacs{03.67.Hk, 03.65.Ud, 42.50.Dv}
\begin{abstract}
We propose two schemes for the generation of the cluster states.
One is based on cavity quantum electrodynamics (QED) techniques.
The scheme only requires resonant interactions between two atoms
and a single-mode cavity. The interaction time is very short,
which is important in view of decoherence. Furthermore, we also
discuss the cavity decay and atomic spontaneous emission case. The
other is based on atomic ensembles. The scheme has inherent fault
tolerance function and is robust to realistic noise and
imperfections. All the facilities used in our schemes are well
within the current technology.
\end{abstract}

\maketitle

\section{introduction}
In the realm of quantum information, entanglement is a universal
resource. Some striking applications of entanglement have been
proposed, such as quantum dense coding \cite{dense}, quantum
teleportation \cite{teleport}, quantum cryptography \cite{ekert},
etc. Generally, entangled states are used as a medium to transfer
quantum information in quantum communication protocols. Moreover,
they are used to speed up computation in quantum algorithms. While
bipartite entanglement is well understood, multipartite
entanglement is still under extensive exploration. For tripartite
entangled quantum system, it falls into two classes of irreducible
entanglement \cite{trientangle,Vidal,Bru}. Recently, Briegel
\emph{et al}. \cite{briegel} introduced a class of \emph{N}-qubit
entangled states, \emph{i.e}., the cluster states, which have some
special properties. The cluster states share the properties both
of the GHZ- and of W-class entangled states. But they still have
some unique properties, \emph{e g}., they have a large persistency
of entanglement, that is, they (in the case of $N>4$) are harder
to be destroyed by local operations than GHZ-class states. In
addition, they can be regarded as a resource for other multi-qubit
entangled states. Thus the cluster states become an important
resource in many branches of physics, especially in quantum
information. Therefore, a number of applications using cluster
states in quantum computation have been proposed \cite{1,2,3,4}.

The generation of the cluster states attracted much attention.
Recently Zou \emph{et al}. proposed probabilistic schemes for
generating the cluster states of four distant trapped atoms in
leaky cavities \cite{zou1}, generating the cluster states in
resonant microwave cavities \cite{zou2} and generating the cluster
states in linear optics system \cite{zou3}. Barrett \emph{et al}.
proposed a protocol for generation of the cluster states using
spatially separated matter qubits and single-photon interference
effects \cite{Barrett} and so on \cite{A,B}.

On the other hand, cavity quantum electrodynamics (QED) technique
is a promising candidate for realizing the quantum processors.
Meanwhile,  much attention was paid to atomic ensembles in
realizing the scalable long-distance quantum communication
\cite{duan1}. The schemes based on atomic ensembles have some
peculiar advantages compared with the schemes of quantum
information processing by the control of single particles.
Firstly, the schemes have inherent fault tolerance function and
are robust to realistic noise and imperfections. Laser
manipulation of atomic ensembles without separately addressing the
individual atoms is dominantly easier than the coherent control of
single particles. In addition, atomic ensembles with suitable
level structure could have some kinds of collectively enhanced
coupling to certain optical mode due to the multi-atom
interference effects. Due to the above distinct advantages, a lot
of novel schemes for the generation of quantum entangled states
and quantum information processing have been proposed by using
atomic ensembles \cite{duan2,duan3,xue,lukin,liu}. Thus in this
paper, we propose two schemes for the generation of the cluster
states using cavity QED technique and the atomic ensembles. Our
cavity QED scheme is different from that in Refs \cite{zou1,zou2}.
The scheme only requires resonant interactions between two atoms
and a single-mode cavity. The interaction time is very short,
which is important in view of decoherence. More important, we
consider the cavity decay and atomic spontaneous emission, which
is unavoidable in the real process of generation. The proposal can
be used to realize logic gates and directly transfer quantum
information from one atom to another one \cite{Zheng} without
using the cavity mode as the memory required in the previous
experiment of Ref \cite{X}. The scheme is very simple and can be
generalized to the ion trap system. But for atomic ensembles
scheme, as far as we known, this is the first scheme for the
generation of the cluster states.

The paper is organized as follows: In section II , we introduce
the cavity-QED model for generating a two-atom cluster state with
and without cavity decay and atomic spontaneous emission, and then
extend the scheme for two-atom cluster state to multi-atom cluster
states case. Necessary discussions are also given in the end of
the section. In section III, we discuss the scheme for generating
the cluster states via atomic ensembles and then conclude the
section and discuss the feasibility of our scheme. The conclusions
appear in section IV.

\section{generation of the cluster states with resonant interactions}
\begin{figure}[tbp]
\includegraphics[scale=0.15,angle=0]{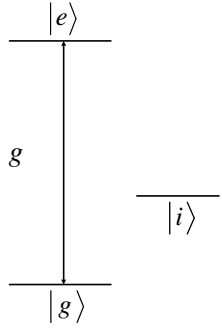}
\caption{The level structure of the atoms. $|g\rangle$ is the ground
state, $|e\rangle$ is the excited state. The cavity mode is
resonantly coupled to the $|e\rangle\leftrightarrow |g\rangle$
transition. The third level $|i\rangle$ is not affected by the
interaction.} \label{fig1}
\end{figure}

In this section , we first use the resonant interaction between
two atoms and a single-mode cavity to generate a two-atom cluster
state. Three-level atoms are used in this model. The relevant
atomic level structure is shown in Fig. \ref{fig1}. The third
level $|i\rangle$ is not affected during the atom-cavity resonant
interaction. Thus the Hamiltonian of the atom-cavity interaction
can be expressed as, in the interaction picture (assuming
$\hbar=1$)\cite{Zheng}

\begin{equation}
\label{1}
H=g_{1}(a^{+}S_{1}^{-}+aS_{1}^{+})+g_{2}(a^{+}S_{2}^{-}+aS_{2}^{+}),
\end{equation}
where $g_{1}$ and $g_{2}$ are the coupling strength of the atoms
1, 2 with the cavity, respectively. $S^{+}=|e\rangle\langle g|$,
$S^{-}=|g\rangle\langle e|$ and $|g\rangle$ is the ground state of
the atoms, $|e\rangle$ is the excited state of the atoms. $a^{+}$,
$a$ are the creation and annihilation operators for the cavity
mode. Assume that the cavity mode is initially prepared in the
vacuum state $|0\rangle$. In order to generate a two-atom cluster
state, we prepare atom 1 in the state
$|\phi\rangle_{1}=\frac{1}{\sqrt{2}}(|g\rangle_{1}+|e\rangle_{1})$
and atom 2 in the state
$|\phi\rangle_{2}=\frac{1}{\sqrt{2}}(|g\rangle_{2}+|i\rangle_{2})$.
So the initial state of the system is
\begin{equation}
\label{2}
|\phi\rangle_{12v}=\frac{1}{2}(|g\rangle_{1}+|e\rangle_{1})\otimes(|g\rangle_{2}+|i\rangle_{2})\otimes|0\rangle.
\end{equation}

Then we send the two atoms through the vacuum cavity, we can
obtain the evolution \cite{Zheng}
\begin{subequations}
\label{3}
\begin{eqnarray}
|eg\rangle_{12}|0\rangle\rightarrow\frac{g_{1}}{E}[\frac{1}{E}(g_{1}cos(Et)+\frac{g_{2}^{2}}{g_{1}})|eg\rangle_{12}|0\rangle\nonumber\\
+\frac{1}{E}g_{2}[cos(Et)-1]|ge\rangle_{12}|0\rangle-i
sin(Et)|gg\rangle_{12}|1\rangle],
\end{eqnarray}
\begin{equation}
|ei\rangle_{12}|0\rangle\rightarrow[cos(g_{1}t)|e\rangle_{1}|0\rangle-i
sin(g_{1}t)|g\rangle_{1}|1\rangle]|i\rangle_{2},
\end{equation}
\begin{equation}
|gg\rangle_{12}|0\rangle\rightarrow|gg\rangle_{12}|0\rangle,
\end{equation}
\begin{equation}
|gi\rangle_{12}|0\rangle\rightarrow|gi\rangle_{12}|0\rangle,
\end{equation}
\end{subequations}
where $E=\sqrt{g_{1}^{2}+g_{2}^{2}}$. If we choose
\begin{equation}
\label{4} t=\frac{\pi}{g_{1}},  g_{2}=\sqrt{3}g_{1},
\end{equation}
which can be achieved by choosing coupling strengths and
interaction time appropriately. Thus, we have
\begin{subequations}
\label{5}
\begin{equation}
|eg\rangle_{12}|0\rangle\rightarrow |eg\rangle_{12}|0\rangle,
\end{equation}
\begin{equation}
|ei\rangle_{12}|0\rangle\rightarrow -|ei\rangle_{12}|0\rangle,
\end{equation}
\begin{equation}
|gg\rangle_{12}|0\rangle\rightarrow |gg\rangle_{12}|0\rangle,
\end{equation}
\begin{equation}
|gi\rangle_{12}|0\rangle\rightarrow |gi\rangle_{12}|0\rangle.
\end{equation}
\end{subequations}
Then send atom 2 through a classical field tuned to the transition
\begin{equation}
\label{6} |i\rangle_{2}\rightarrow -|e\rangle_{2}.
\end{equation}
These lead the state of atoms 1 and 2 to
\begin{eqnarray}
\label{7}
|\phi\rangle_{12}&=&\frac{1}{2}[|g\rangle_{1}(|g\rangle_{2}-|e\rangle_{2})+|e\rangle_{1}(|g\rangle_{2}+|e\rangle_{2})]\nonumber\\
&=&\frac{1}{2}(|g\rangle_{1}\sigma_{z}^{2}+|e\rangle_{1})(|g\rangle_{2}+|e\rangle_{2}).
\end{eqnarray}
Obviously we get a standard two-atom cluster state. While in the
real processing, the cavity decay and atomic spontaneous emission
are unavoidable. Thus the discussion of these is necessary. Taking
the cavity decay and atomic spontaneous emission into
consideration, the Hamiltonian of the atom-cavity interaction can
be expressed as (under the condition that no photon is detected
either by the spontaneous emission or by the leakage of a photon
through the cavity mirror and assuming $\hbar=1$)
\begin{eqnarray}
\label{8}
H&=&g_{1}(a^{+}S_{1}^{-}+aS_{1}^{+})+g_{2}(a^{+}S_{2}^{-}+aS_{2}^{+})\nonumber\\&-&
i\frac{\kappa}{2}a^{+}a -
i\frac{\tau}{2}\Sigma_{j=1}^{2}|e\rangle_{j}\langle e|,
\end{eqnarray}
where $\kappa$ is the cavity decay rate and $\tau$ is the atomic
spontaneous emission rate. If we send the atoms 1 and 2 through
the vacuum cavity, choose the coupling strengths, interaction time
$g_{2}=\sqrt{3}g_{1}$, $t=\frac{\pi}{g_{1}}$ appropriately and set
$\kappa=\tau=0.1g_{1}$, and then send the atom 2 through a
classical field as in Eq. (\ref{6}).  Thus the state of the atoms
1 and 2  becomes
\begin{eqnarray}
\label{9}
|\phi\rangle_{12}=\sqrt{\frac{1}{2(1+e^{-\pi/10})}}[|g\rangle_{1}(|g\rangle_{2}-|e\rangle_{2})
\nonumber\\+e^{-\pi/20}|e\rangle_{1}(|g\rangle_{2}+|e\rangle_{2})].
\end{eqnarray}
The fidelity of this state relative to the standard two-atom
cluster state in Eq. (7) is
$\frac{(1+e^{-\pi/20})^{2}}{2({1+e^{-\pi/10}})}\simeq0.994$ and
the probability of success is
$\frac{1+e^{-\pi/10}}{2}\simeq0.865$. The fidelity and probability
approach perfection.

Multi-atom entanglement is a very important source in quantum
information processing and quantum computation. Especially the
multi-atom cluster states attract many scientific attention
recently, and some of their applications have been proposed
\cite{Tame, Zhou, Raussendorf, Nielsen}. Thus the generation of
multi-atom cluster states is vital for the construction of the
practical quantum computers. Here, we generalize the above scheme
of two-atom cluster state to multi-atom cluster states case.

\begin{figure}[tbp]
\includegraphics[scale=0.35,angle=0]{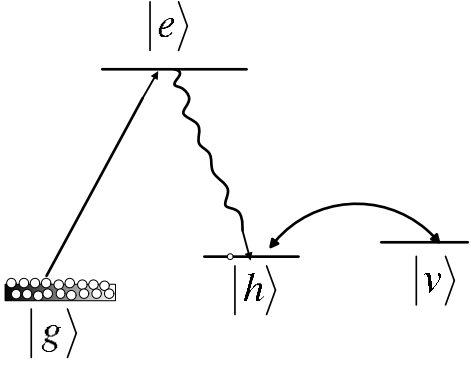}
\caption{The relevant atomic level structure of alkali metal atom.
The transition of $|e\rangle\rightarrow|h\rangle$ can emit a
forward-scattered Stokes photon co-propagating with the laser pulse.
The excitation in the mode $h$ can be transferred to optical
excitation by applying an anti-pump pulse.} \label{fig2}
\end{figure}

We first prepare $N$ ($N\geq 2$) atoms in the states
\begin{subequations}
\begin{equation}
\label{10}
|\phi\rangle_{1}=\frac{1}{\sqrt{2}}(|g\rangle_{1}+|e\rangle_{1}),
\end{equation}
\begin{equation}
|\phi\rangle_{j}=\frac{1}{\sqrt{2}}(|g\rangle_{j}+|i\rangle_{j}),
\end{equation}
\end{subequations}
where $j=2,3\cdots N$. The $N-1$ cavities are all prepared in
vacuum states $|0\rangle$. So the total state of atoms is
\begin{equation}
\label{11}
|\phi\rangle_{1j}=\frac{1}{2^{N/2}}(|g\rangle_{1}+|e\rangle_{1})\bigotimes_{j=2}^{N}(|g\rangle_{j}+|i\rangle_{j}).
\end{equation}

For the case of ideal cavity, firstly, we send atoms 1 and 2
through a vacuum cavity. The interaction between atoms 1, 2 and
the cavity mode is governed by the Hamiltonian of Eq. (\ref{1}).
Meanwhile, we choose the coupling strengths, interaction time
appropriately as in Eq. (\ref{4}). Then send atom 2 through a
classical field as in Eq. (\ref{6}). These lead Eq. (\ref{11}) to
\begin{equation}
\label{12}
|\phi\rangle_{1j}=\frac{1}{2^{N/2}}(|g\rangle_{1}\sigma_{z}^{2}+|e\rangle_{1})
(|g\rangle_{2}+|e\rangle_{2})\bigotimes_{j=3}^{N}(|g\rangle_{j}+|i\rangle_{j}).
\end{equation}
Next, we send atoms 2 and 3 through another vacuum cavity. After
the same interaction as on atoms 1 and 2, send atom 3 through a
classical field as in Eq. (\ref{6}), Here, Eq. (\ref{12}) becomes
\begin{eqnarray}
\label{13}
|\phi\rangle_{1j}=\frac{1}{2^{N/2}}(|g\rangle_{1}\sigma_{z}^{2}+|e\rangle_{1})
(|g\rangle_{2}\sigma_{z}^{3}+|e\rangle_{2})\nonumber\\
(|g\rangle_{3}+|e\rangle_{3})\bigotimes_{j=4}^{N}(|g\rangle_{j}+|i\rangle_{j}).
\end{eqnarray}
From the form of  above states, we can conclude if we send two atoms
through a vacuum cavity every time and then send one (the bigger
subscript) of the two atoms through a classical filed, step by step,
we can obtain the multi-atom cluster states easily. In other words,
firstly we send atoms 1 and 2 through a vacuum cavity, then send
atom 2 through a classical field. Secondly, we send atoms 2 and 3
through another vacuum cavity, then send atom 3 through another
classical field, $\cdots$. Finally, we send atoms $N-1$ and $N$
through the last vacuum cavity, then send atom $N$ through a
classical field. Thus the multi-atom cluster states can be obtained
\begin{equation}
\label{14} |\phi\rangle_{N}=\frac{1}{2^{N/2}} \bigotimes_{j=1}
^{N}(|g\rangle_{j}\sigma _{z}^{j+1}+|e\rangle_{j}),
\end{equation}
where $\sigma _{z}^{N+1}\equiv1$.

For the case of real processing (with cavity decay and atomic
spontaneous emission), we can obtain the cluster state by the same
process as in the above ideal case and set $\kappa=\tau=0.1g_{1}$.
We can obtain the cluster states
\begin{eqnarray}
\label{15}
|\phi\rangle_{N}&=&\sqrt{\frac{1}{2(1+e^{-\pi/10})^{N-1}}}\bigotimes_{j=1}
^{N-1}(|g\rangle_{j}\sigma
_{z}^{j+1}+e^{-\pi/20}|e\rangle_{j})\nonumber\\
&\otimes&(|g\rangle_{N}\sigma _{z}^{N+1}+|e\rangle_{N}).
\end{eqnarray}
While the fidelity of this state relative to the standard
multi-atom cluster state in Eq. (14) is
$[\frac{(1+e^{-\pi/20})^{2}}{2({1+e^{-\pi/10}})}]^{N-1}$ and the
successful probability of obtaining the multi-atom cluster state
is $(\frac{1+e^{-\pi/10}}{2})^{N-1}$. It is shown that the
successful probability and fidelity both decrease exponentially
with the increase of $N$.

Next, we briefly consider the feasibility of the current scheme. The
scheme requires two atoms in a vacuum cavity, which have different
coupling strengths with the cavity mode. The coupling depends on the
atomic positions: $g=\Omega e^{-r^{2}/\omega^{2}}$, where $\Omega$
is the coupling strength at the cavity center, $\omega$ is the waist
of the cavity mode, and $r$ is the distance between the atom and the
cavity center \cite{p}. The condition $g_{2}=\sqrt{3}g_{1}$ in our
scheme can be satisfied by locating one atom at the center of the
cavity and locating the other one at the position $r=\omega
\sqrt{ln\sqrt{3}}$. According to the recent experiments with Cs
atoms trapped in an optical cavity\cite{v}, the condition can be
obtained.

For the resonant cavity, in order to generate the cluster states
successfully, the relationship between the interaction time and the
excited atom lifetime should be taken into consideration. The
interaction time should be much shorter than that of atom radiation.
Hence, atom with a sufficiently long excited lifetime should be
chosen. For Rydberg atom with principal quantum numbers $50$ and
$51$, the radiative time is $T_{1}\simeq3\times10^{-2}$s. From the
analysis in Ref \cite{biao}, the interaction time is on the order
$T\simeq2\times10^{-4}$s, which is much shorter than the atomic
radiative time. So the condition can be satisfied by choosing
Rydberg atom. Our scheme requires that two atoms be simultaneously
sent through a cavity, otherwise there will be an error. Assume that
during the generation of a two-atom cluster state, one atom enters
the cavity $0.01t$ sooner than another atom, with $t$ being the time
of each atom staying in the cavity. We can obtain the fidelity
$F\simeq 0.999$ for generation of two-atom cluster state. Obviously
in this case the operation is only slightly affected.

Furthermore one needs to reach the Lamb-Dicke regime in order to
generate the cluster states successfully. For the initial state of
Eq. (\ref{2}), in the Lamb-Dicke regime, the infidelity caused by
the spatial extension of the atomic wave function is about
$\Delta\simeq (ka)^{2}\pi$, where $k$ is the wave vector of the
cavity mode and $a$ is the spread of the atomic wave function.
Setting $\Delta\simeq0.01$, we have $a\simeq0.01\lambda$, where
$\lambda$ is the wavelength of the cavity mode. If the atom
trajectories cross the cavity with the deviation of less $0.1$
degree from its pre-determined direction, we can ensure the fidelity
is about $0.999$ for generation of two-atom cluster state. While in
order to maintain $g_{2}=\sqrt{3}g_{1}$ in the process of atomic
motion in the cavity, we can choose the parameter of cavity
$z\leq0.5z_{0}$, where $z_{0}=\frac{\pi\omega^{2}}{\lambda}$ and
$2z$ is the length of the cavity. We can obtain the error is only
about $10^{-3}$. In the these cases, we can obtain the fidelity
$F\simeq 0.999$ for generation of two-atom cluster state, which is
bigger than the case of cavity decay and atomic spontaneous emission
in the process of generation. Therefore our scheme is feasible with
the current cavity QED technology.

The scheme for generating the cluster states in cavity QED only
requires resonant interactions between two atoms and a single-cavity
mode. The interaction time is very short, which is very important in
view of decoherence. For the ideal case, the successful probability
and the fidelity are both perfect (equal to $1.0$). For the real
case, the successful probability is $0.865$ and the fidelity is
$0.994$ for the two-atom cluster states, while the successful
probability and the fidelity for the multi-atom cluster states both
decrease exponentially with the increase of $N$. The scheme is very
simple and can be generalized to the ion trap system.

\section{generation of the cluster states with atomic ensembles}
In this section, we first introduce the basic system using in this
paper. Atomic ensembles consist of  a large number of identical
alkali metal atoms. The relevant level structure of the alkali
metal atoms is shown in Fig. \ref {fig2}. $|g\rangle$ is the
ground state, $|e\rangle$ is the excited state and $|h\rangle$,
$|v\rangle$ are two metastable states for storing a qubit of
information, \emph{e g.}, Zeeman or hyperfine sublevels. For the
three levels $|g\rangle$, $|h\rangle $ and $|v\rangle $, which can
be coupled via a Raman process, two collective atomic operators
can be defined as
$$s=(1/\sqrt{N_{a}})\Sigma_{i=1}^{N_{a}}|g\rangle_{i}\langle s|,$$
where $s = h,v$, and $N_{a}\gg1$ is the total number of atoms. $s$
are similar to independent bosonic mode operators provided that all
the atoms remain in ground state $|g\rangle$. The states of the
atomic ensemble can be expressed as $|s\rangle = s^{+}|vac\rangle$
($s = h,v$) after the emission of the single Stokes photon in a
forward direction, where $|vac \rangle \equiv \otimes
_{i=1}^{N_{a}}|g\rangle _{i}$ denotes the ground state of the atomic
ensemble.

It is necessary to discuss the realization of Controlled-Not gate
for the generation of cluster states. The Controlled-Not gate can be
realized via atomic ensembles with the help of Raman laser
manipulations, beam splitters, and single-photon detections.
Realization of Bell-basis measurement and generation of tripartite
GHZ states is important for realization of Controlled-Not gate.
Bell-basis measurement can be realized using the setup in Fig.
\ref{fig2}. The four Bell states of the system are $|\phi\rangle
_{AB}^\pm = ( h_{A}^ {+} h_{B}^{+} \pm v_{A}^ {+} v_{B}^
{+})|vac\rangle _{AB}/\sqrt{2}$
 and $|\varphi\rangle _{AB}^\pm = ( h_{A}^
{+} v_{B}^{+} \pm v_{A}^ {+}h_{B}^ {+})|vac\rangle _{AB}/\sqrt{2}$.
We can use the setup to achieve the task, as shown in Fig.
\ref{fig2}. Firstly, we apply anti-pump laser pulses to the two
atomic ensembles $ A $ and $ B$ to transfer their $h$ excitations to
optical excitations, and detect the anti-Stokes photons by detectors
$D$1 and $D$2. If only detector $D$1 (or $D$2) clicks, we will apply
single-qubit rotations to both ensembles to rotate their $v$ modes
to $h$ modes by shinning $\pi$ length Raman pulses or
radio-frequency pulses on the two ensembles $A$ and $B$. Then we
apply anti-pump laser pulses to two atomic ensembles $A$ and $B$
again, and detect anti-Stokes photons by $D$1 and $D$2. Now, there
are two different results of detection: (1) If detector $D$1 (or
$D$2) clicks (one detector clicks twice in the two detections),
post-select the cases that each ensemble has only one excitation,
atomic ensembles $A$ and $B$ are projected into $|\varphi\rangle
_{AB}^{+}= ( h_{A}^ {+} v_{B}^{+}+ v_{A}^ {+}h_{B}^ {+})|vac\rangle
_{AB}/\sqrt{2}$; (2)If $D$2 (or $D$1) clicks (detectors $D$1 and
$D$2 click respectively in the two detections), post-select the
cases that each ensemble has only one excitation, atomic ensembles
$A$ and $B$ are projected into $|\varphi\rangle _{AB}^-= ( h_{A}^
{+} v_{B}^{+}- v_{A}^ {+}h_{B}^ {+})|vac\rangle _{AB}/\sqrt{2}$.
Obviously, if we add single-qubit rotations in the above process, we
can realize the projection of $|\phi\rangle _{AB}^\pm = ( h_{A}^ {+}
h_{B}^{+} \pm v_{A}^ {+}v_{B}^ {+})|vac\rangle _{AB}/\sqrt{2}$ by
post-selecting sense.

Tripartite GHZ states can be prepared using the protocol of Ref
\cite{duan2} with atomic ensembles.  Firstly, atomic ensembles 1 and
2 can be prepared in the state
$|\phi\rangle_{12}^{\pm}=(h_{1}^{+}\pm
e^{i\varphi}h_{2}^{+})|vac\rangle_{12}/\sqrt{2}$ as in Ref
\cite{duan1}. Then we can omit $e^{i\varphi}$ by the way in
\cite{duan2} and perform a single-qubit rotation on atomic ensemble
2. The state of atomic ensembles 1 and 2 becomes
$|\phi\rangle_{12}=(h_{1}^{+}+v_{2}^{+})|vac\rangle_{12}/\sqrt{2}$.
Secondly, we prepare the atomic ensembles 2, 3 and 3, 1 in the
states
$|\phi\rangle_{23}=(h_{2}^{+}+v_{3}^{+})|vac\rangle_{23}/\sqrt{2}$
and
$|\phi\rangle_{31}=(h_{3}^{+}+v_{1}^{+})|vac\rangle_{31}/\sqrt{2}$.
So the total state becomes
$|\phi\rangle_{123}=|\phi\rangle_{12}\otimes|\phi\rangle_{23}\otimes|\phi\rangle_{31}$.
Post-select the case that each ensemble has only one excitation, we
can obtain the GHZ state
$|\phi\rangle_{123}=(h_{1}^{+}h_{2}^{+}h_{3}^{+}+v_{1}^{+}v_{2}^{+}v_{3}^{+})|vac\rangle_{123}/\sqrt{2}$.
In the same way, we can prepare another GHZ state using atomic
ensembles 4, 5 and 6 $|\phi\rangle _{456} = (h_{4}^ {+} h_{5}^{ +}
h_{6}^ {+} + v_{4}^ {+} v_{5}^{+} v_{6}^{+})|vac\rangle_{456}/{\sqrt
2 }$.

In order to realize C-NOT gate, we prepare two atomic ensembles 7
and 8 ( ensemble 7 as control, ensemble 8 as target), which are in
$|\phi \rangle _{7} = ( h_{7}^{+} + v_{7}^{+})|vac\rangle_{7}$
 and $|\phi\rangle _{8} =(h_{8}^{+} - v_{8}^ {+} )|vac
\rangle _{8}$ by single-qubit rotations. Firstly, we apply Hadamard
transformations on atomic ensembles 1, 2 and 3 respectively, and
then make a Bell-basis measurement on atomic ensembles 3 and 4. Then
the state $|\phi\rangle _{123456}$ collapses to one of the following
four unnormalized states
\begin{subequations}
\begin{eqnarray}
 |\phi\rangle_{1256}&=&[(h_{1}^{+}h_{2}^{+} +
v_{1}^{+}v_{2}^ {+})h_{5}^{+} h_{6}^{+}\nonumber\\
&\pm&(h_{1}^{+}v_{2}^{+}+v_{1}^{+}h_{2}^{+})v_{5}^{+}v_{6}^{+}]|vac\rangle_{1256},
\end{eqnarray}
\begin{eqnarray}
|\varphi\rangle_{1256}&=&[(h_{1}^{+}h_{2}^{+}+v_{1}^{+}v_{2}^{+})v_{5}^{+}v_{6}^{+}\nonumber\\
&\pm&(h_{1}^{+}v_{2}^{+}+v_{1}^{+}h_{2}^{+})h_{5}^{+}h_{6}^{+}]|vac\rangle_{1256}.
\end{eqnarray}
\end{subequations}
Where $|\phi\rangle_{1256}$ and $|\varphi\rangle _{1256}$ are the
results of the projection into $|\phi\rangle _{34}^\pm $ and
$|\varphi \rangle _{34}^\pm $, respectively. They can unify as
$|\chi\rangle_{1256}=[(h_{1}^{+}h_{2}^{+} + v_{1}^{+}v_{2}^
{+})h_{5}^{+} h_{6}^{+}+(h_{1}^{+}v_{2}^{+}
+v_{1}^{+}h_{2}^{+})v_{5}^{+}v_{6}^{+}]|vac\rangle_{1256}$ with the
help of simple single-qubit operations.

Then we make Bell-basis measurements on atomic ensembles 1, 8 and 6,
7. The state of atomic ensembles 2 and 5 collapses to one of the
following states
\begin{subequations}
\begin{equation}
\label{25a}|\phi\rangle_{25}=(h_2^{+} - v_2^{+})(h_5^{+}
-v_5^{+})|vac\rangle_{25}/2,
\end{equation}
\begin{equation}
\label{25b}|\varphi\rangle_{25}=(h_2^{+} - v_2^{+})(h_5^{+}
+v_5^{+})|vac\rangle_{25}/2.
\end{equation}
\end{subequations}
where Eq. (\ref{25a}) corresponds to the measurement results of
$|\phi \rangle _{67}^{+}$ and $|\varphi\rangle_{67}^{+}$, and Eq.
(\ref{25b}) corresponds to $|\phi\rangle_{67}^{-} $ and $|\varphi
\rangle_{67}^{-}$. We can transform state (\ref{25b}) to state
(\ref{25a}) by single-qubit rotations. Obviously, C-NOT gate has
been realized and the state of atomic ensembles 7 and 8 has been
mapped on ensembles 2 and 5.

\begin{figure}[tbp]
\includegraphics[scale=0.35,angle=90]{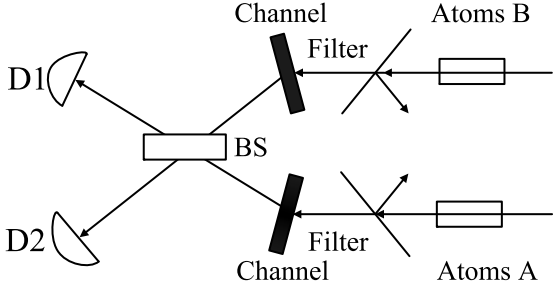}
\caption{Setup of realizing Bell-basis measurement. The two atomic
ensembles A and B are pencil-shaped, which are illuminated by the
synchronized laser pulses. The forward-scattered anti-Stokes photons
are collected and coupled to optical channel (fiber) after the
filter. BS is a 50/50 beam splitter, and the outputs are detected by
two single-photon detectors $D$1 and $D$2.} \label{fig2}
\end{figure}

Next, we discuss the generation of bipartite cluster state. The
atomic ensembles 1 and 2 are initially prepared in the state
\begin{equation}
\label{16} |\phi\rangle_{12} =v^{+}_{1}v^{+}_{2}|vac\rangle_{12}
\end{equation}
using Raman pulses. All the single-qubit transformation can be
achieved by laser pulses in atomic ensembles. Secondly, we perform
a single-qubit operation on atomic ensemble 1

\begin{equation}
\label{17} v^{+}_{1}|vac\rangle_{1} \rightarrow (h^{+}_{1}+
v^{+}_{1})|vac\rangle_{1}/\sqrt{2}.
\end{equation}

Then, we perform a Controlled-Not transformation on the two atomic
ensembles, where atomic ensemble 1 serving as control qubit and
atomic ensemble  2 as target qubit.  Now, the above procedures lead
Eq. (\ref{16}) to

\begin{equation}
\label{18} |\phi\rangle_{12} = (h^{+}_{1}v^{+}_{2}+
v^{+}_{1}h^{+}_{2})|vac\rangle_{12}/\sqrt{2}.
\end{equation}

Finally, we perform a single-qubit operation on atomic ensemble 1

\begin{eqnarray}
\label{19} h^{+}_{1}|vac\rangle_{1} \rightarrow
v^{+}_{1}|vac\rangle_{1}, v^{+}_{1}|vac\rangle_{1} \rightarrow
h^{+}_{1}|vac\rangle_{1},
\end{eqnarray}
and another single-qubit operation on atomic ensemble 2

\begin{eqnarray}
\label{20}h^{+}_{2}|vac\rangle_{2} \rightarrow (h^{+}_{2}-
v^{+}_{2})|vac\rangle_{2}/\sqrt{2},\nonumber\\v^{+}_{2}|vac\rangle_{2}
\rightarrow (h^{+}_{2}+ v^{+}_{2})|vac\rangle_{2}/\sqrt{2}.
\end{eqnarray}
Here, the quantum state of atomic ensembles 1 and 2 becomes

\begin{eqnarray}
\label{21} |\phi\rangle_{12} &=& [h^{+}_{1}(h^{+}_{2}-v^{+}_{2})+
v^{+}_{1}(h^{+}_{2}+v^{+}_{2})]|vac\rangle_{12}/2\nonumber\\&=&
[(h^{+}_{1}\sigma_{z}^{2}+v^{+}_{1})
(h^{+}_{2}+v^{+}_{2})]|vac\rangle_{12}/2.
\end{eqnarray}

Obviously the state is a standard bipartite cluster states
($N=2$). The cluster states ($N=2, 3$) can be also generated
without Controlled-Not transformation \cite{duan1, duan2}.
However, for the generation of the multipartite cluster states,
using the proposals of  Ref. \cite{duan1, duan2} are very hard,
while, it can be realized by the above method with Controlled-Not
transformations, as shown below.

Out of question, for the generation of arbitrary $N$-particle
cluster state ($N\geq 2$), we can use the single-qubit operations
and controlled-not transformations to achieve the task perfectly.
Here, we discuss the process in detail. Firstly, we prepare $N$
atomic ensembles, which are all in the states
$v^{+}_{i}|vac\rangle_{i}$ ($i=1,2\cdots N$). So the state of the
whole system is
\begin{equation}
\label{22} |\phi\rangle_{12\cdots N} = (v^{+}_{1}v^{+}_{2}\cdots
v^{+}_{N})|vac\rangle_{12\cdots N}.
\end{equation}
Secondly, we perform appropriately transformations as the above
process on atomic ensembles 1 and 2 (Eq.(\ref{17})-(\ref{20})),
which lead the initial state to
\begin{eqnarray}
\label{23} |\phi\rangle_{12\cdots N}=
(h^{+}_{1}\sigma_{z}^{2}+v^{+}_{1}) (h^{+}_{2}+v^{+}_{2})
\nonumber\\(v^{+}_{3}v^{+}_{4}\cdots
v^{+}_{N})|vac\rangle_{12\cdots N}/2.
\end{eqnarray}
Then, we perform the same transformations on atomic ensembles 2
and 3 as atomic ensembles 1 and 2. We can obtain the result
\begin{eqnarray}
\label{24} |\phi\rangle_{12\cdots N} =
(h^{+}_{1}\sigma_{z}^{2}+v^{+}_{1})
(h^{+}_{2}\sigma_{z}^{3}+v^{+}_{2})(h^{+}_{3}+v^{+}_{3})\nonumber\\
(v^{+}_{4}v^{+}_{5}\cdots v^{+}_{N})|vac\rangle_{12\cdots
N}/2\sqrt{2}.
\end{eqnarray}

In a word, if we perform the transformations of Eq.
(\ref{17})-(\ref{20}) on atomic ensembles 1 and 2, then on atomic
ensembles 2 and 3, up to on atomic ensembles $N-1$ and $N$, we
will obtain the perfect multipartite cluster states

\begin{eqnarray}
\label{25} |\phi\rangle_{12\cdots N}& =&
\frac{1}{2^{N/2}}(h^{+}_{1}\sigma_{z}^{2}+v^{+}_{1})
(h^{+}_{2}\sigma_{z}^{3}+v^{+}_{2})\nonumber\\ &\cdots&
(h^{+}_{N}+v^{+}_{N}) |vac\rangle_{12\cdots N}\nonumber\\&=&
\frac{1}{2^{N/2}}
\bigotimes^{N}_{i=1}(h^{+}_{i}\sigma_{z}^{i+1}+v^{+}_{i})|vac\rangle_{12\cdots
i},
\end{eqnarray}
where $\sigma_{z}^{N+1}\equiv1$.

We briefly discuss the feasibility of the current scheme. If we want
to generate a high-fidelity entangled state about $16$ ensembles, a
time $T_{imp}\simeq 50 ms$ will be needed by choosing other
parameters appropriately, which has been proved \cite{duan2}. With
such a short preparation time $T_{imp}$, the noise that we have not
included is negligible, such as the nonstationary phase drift
induced by the pumping phase or by the optical channel. As long as
the number $n$ of the ensembles is not huge, we also can safely
neglect the single-bit rotation error ( below $10^{-4}$ with the use
of accurate polarization techniques for Zeeman sublevels \cite{gg} )
and the dark count probability of single-photon detectors (about
$10^{-5}$ in a typical detection time window $0.1$ $\mu s$
\cite{duan2}). Thus it seems reasonable to generate cluster states
over tens of ensembles with the current technology.  Furthermore,
the scaling can be made polynomial by dividing the whole preparation
process into small steps, checking in each steps and repeating these
steps instead of the whole process in case it fails. So our scheme
has inherent fault tolerance function and is robust to realistic
noise and imperfections  \cite{duan1,duan2}.

The physical scheme for generating the cluster states based on
atomic ensembles have some peculiar advantages compared with the
schemes by the control of single particles,\emph{ e g.} the schemes
have inherent fault tolerance function and are robust to realistic
noise and imperfections. Laser manipulation of atomic ensembles
without separately addressing the individual atoms is dominantly
easier than the coherent control of single particles. Atomic
ensembles with suitable level structure could have some kinds of
collectively enhanced coupling to certain optical mode due to the
multi-atom interference effects and so on\cite{duan1}. At the same
time, We can generate the $N$-qubit cluster state simply by
extending the two-qubit case.

\section{conclusions}
We propose two schemes for the generation of the cluster states. One
scheme is based on cavity quantum electrodynamics (QED) technics.
The scheme only requires resonant interactions between two atoms and
a single-mode cavity. The interaction time is very short, which is
important in view of decoherence. We first introduce the two-atom
case then extend it to mult-atom case. Furthermore, we consider the
cavity decay and atomic spontaneous emission case, the successful
probability and the fidelity for the multi-atom cluster states both
decrease exponentially with the increase of $N$. The scheme is very
simple and can be generalized to the ion trap system. The other is
based on atomic ensembles. The scheme has inherent fault tolerance
function and is robust to realistic noise and imperfections. The
generation of cluster states from two-qubit case to multi-qubit case
is simple and feasible. All of the facilities used in our schemes
are well within the current technology.

\begin{acknowledgments}
This work is supported by the Natural Science Foundation of the
Education Department of Anhui Province under Grant No: 2006kj070A
and Anhui Provincial Natural Science Foundation under Grant No:
03042401 and the Talent Foundation of Anhui University.
\end{acknowledgments}

\end{document}